\documentclass[conference]{IEEEtran}

\usepackage[pdftex]{graphicx}
\usepackage{xcolor}
\usepackage[hidelinks]{hyperref}

\usepackage{balance}

\definecolor{Paired-1}{RGB}{31,120,180}
\definecolor{Paired-2}{RGB}{166,206,227}
\definecolor{Paired-3}{RGB}{51,160,44}
\definecolor{Paired-4}{RGB}{178,223,138}
\definecolor{Paired-5}{RGB}{227,26,28}
\definecolor{Paired-6}{RGB}{251,154,153}
\definecolor{Paired-7}{RGB}{255,127,0}
\definecolor{Paired-8}{RGB}{253,191,111}
\definecolor{Paired-9}{RGB}{106,61,154}
\definecolor{Paired-10}{RGB}{202,178,214}
\definecolor{Paired-11}{RGB}{177,89,40}
\definecolor{Paired-12}{RGB}{255,255,153}
\definecolor{Paired-13}{RGB}{80,80,80}
\definecolor{Paired-14}{RGB}{153, 153, 255}

\usepackage{amsmath,amssymb,amsfonts,stmaryrd}
\usepackage{bm}
\usepackage{dsfont}
\usepackage{numprint}
\newcommand{\isep}{\mathrel{{.}\,{.}}\nobreak}

\usepackage{adjustbox}
\usepackage{booktabs}
\usepackage{multirow}

\usepackage[caption=false,font=footnotesize]{subfig}
\usepackage{tikz}
\usetikzlibrary{matrix, positioning, patterns, shapes, arrows}
\usepackage{pgfplots}
\usepackage{amsmath}
 
\usepgfplotslibrary{groupplots}
\usepgfplotslibrary{colorbrewer}
\usetikzlibrary{spy,backgrounds}

\usepackage{pgfplots}
\pgfplotsset{compat=newest}
\usepgfplotslibrary{groupplots}

\usepackage{tikz}
\usetikzlibrary{matrix, positioning, patterns, shapes, arrows}

\usepackage[draft]{fixme} 
\fxsetup{theme=color}

\usepackage{lipsum}

\usepackage[ruled, linesnumbered, vlined]{algorithm2e}

\title{Step-GRAND: A Low Latency Universal Soft-input Decoder  }

\author{\IEEEauthorblockN{ Syed Mohsin Abbas and Chi-Ying Tsui}
\IEEEauthorblockA{\textit{Department of Electronic and Computer Engineering} \\
\textit{The Hong Kong University of Science and Technology}\\
Hong Kong, China\\
Emails: smabbas@connect.ust.hk, eetsui@ust.hk}
\and
\IEEEauthorblockN{ Marwan Jalaleddine and Warren J. Gross }
\IEEEauthorblockA{\textit{Department of Electrical and Computer Engineering} \\
\textit{McGill University}\\
Montr\'eal, Qu\'ebec, Canada \\
marwan.jalaleddine@mail.mcgill.ca, warren.gross@mcgill.ca}
}

\begin{document}

\maketitle

\begin{abstract}

GRAND features both soft-input and hard-input variants that are well suited to efficient hardware implementations that can be characterized with achievable average and worst-case decoding latency. This paper introduces step-GRAND, a soft-input variant of GRAND that, in addition to achieving appealing average decoding latency, also reduces the worst-case decoding latency of the corresponding hardware implementation. The hardware implementation results demonstrate that the proposed step-GRAND can decode CA-polar code $(128,105+11)$ with an average information throughput of $47.7$ Gbps at the target FER of $\leq10^{-7}$. Furthermore, the proposed step-GRAND hardware is $10\times$ more area efficient than the previous soft-input ORBGRAND hardware implementation, and its worst-case latency is $\frac{1}{6.8}\times$ that of the previous ORBGRAND hardware.

\end{abstract}

\begin{IEEEkeywords}
Guessing Random Additive Noise Decoding (GRAND), GRAND with ABandonment (GRANDAB), Ordered Reliability Bits GRAND (ORBGRAND), Maximum Likelihood (ML) decoding, Ultra-Reliable and Low Latency Communications (URLLC). 
\end{IEEEkeywords}

\section{Introduction}
The emergence of novel applications that have stringent requirements for URLLC \cite{3GPP} has recently rekindled a great deal of interest in short channel codes and associated ML decoding approaches. A few examples of these applications include augmented and virtual reality \cite{URLLC2}, intelligent transportation systems \cite{URLLC1}, the internet of things \cite{IoT1} and machine-to-machine communication \cite{URLLC3}.

For short-length and high-rate channel codes, GRAND \cite{Duffy19TIT} has recently been proposed as a universal ML decoding technique. Since GRAND is noise-centric and code-agnostic, it attempts to guess the noise that corrupted the codeword during transmission through the communication channel rather than relying on the structure of the underlying code to decode a codeword. GRAND guesses the noise by generating Test Error Patterns (TEPs), and the order in which these TEPs ($\bm{e}$) are generated is the primary difference between different GRAND variants.

For both hard-input \cite{Duffy19TIT}\cite{Duffy20205g} and soft-input \cite{duffy2020ordered}\cite{solomon2020soft} GRAND variants, several high-throughput and energy-efficient hardware implementations have been reported in the literature \cite{GRANDAB-chip}\cite{GRANDAB-VLSI}\cite{ORBGRAND-TVLSI}\cite{carloORB}\cite{ORBBU}. In general, the decoding latency of GRAND-based hardware implementations can be categorized into average and worst-case decoding latency. While the average decoding latency of GRAND hardware is typically much lower than the worst-case decoding latency, the latter can still pose a significant barrier to the adoption of GRAND based hardware implementations for applications that require strict adherence to both average and worst-case decoding latency requirements, such as the URLLC application scenario \cite{3GPP}\cite{URLLC2}\cite{URLLC1}\cite{IoT1}\cite{URLLC3}.

In this work, we introduce step-GRAND, a soft-input variant of GRAND that not only offers a low average decoding latency but also reduces the worst-case decoding latency compared to soft-input ORBGRAND \cite{ORBGRAND-TVLSI}. We propose a simplified TEP generation scheme and develop a high-throughput VLSI architecture for the proposed step-GRAND. Furthermore, the proposed step-GRAND introduces parameters that can be adjusted to meet the desired decoding performance and complexity/latency constraints for a target application. 

The VLSI implementation results show that for a linear block code with length $128$ $(n)$ and $105$ information bits ($k$), the proposed step-GRAND hardware can achieve an average information throughput of $47.7$ Gbps. Furthermore, in comparison to the previously proposed ORBGRAND hardware implementation \cite{ORBGRAND-TVLSI}, the proposed step-GRAND hardware is $10\times$ more area efficient and the worst-case latency of the step-GRAND hardware is $\frac{1}{6.8}\times$ the worst-case latency of previous ORBGRAND hardware \cite{ORBGRAND-TVLSI}.

The remainder of this paper is structured as follows: Preliminaries on GRAND are provided in Section II. Section III introduces the proposed step-GRAND. The numerical simulation results are presented in Section IV. The proposed step-GRAND VLSI architecture and its implementation results are presented in Section V. Finally, in Section VI, concluding remarks are made.

\section{Preliminaries}
\subsection{Notations}
Matrices are denoted by a bold upper-case letter ($\bm{M}$), while vectors are denoted with bold lower-case letters ($\bm{v}$). The transpose operator is represented by $^\top$. The $i^{\text{th}}$ element of a vector $\bm{v}$ is denoted by $v_i$. The number of $k$-combinations from a given set of $n$ elements is noted by $\binom{n}{k}$. $\mathds{1}_n$ is the indicator vector where all locations except the $n^{\text{th}}$ location are $0$ and the $n^{\text{th}}$ location is $1$. Similarly, $\mathds{1}_{i,j\ldots,z}=\mathds{1}_i\oplus\mathds{1}_j\oplus\ldots\mathds{1}_z$, with $i\neq j \ldots\neq k$. All the indices start at $1$. For this work, all operations are restricted to the Galois field with 2 elements, noted $\mathbb{F}_2$. Furthermore, we restrict ourselves to $(n,k)$ linear block codes.

\begin{figure}[!tb]
  \centering
  \includegraphics[width=1\linewidth]{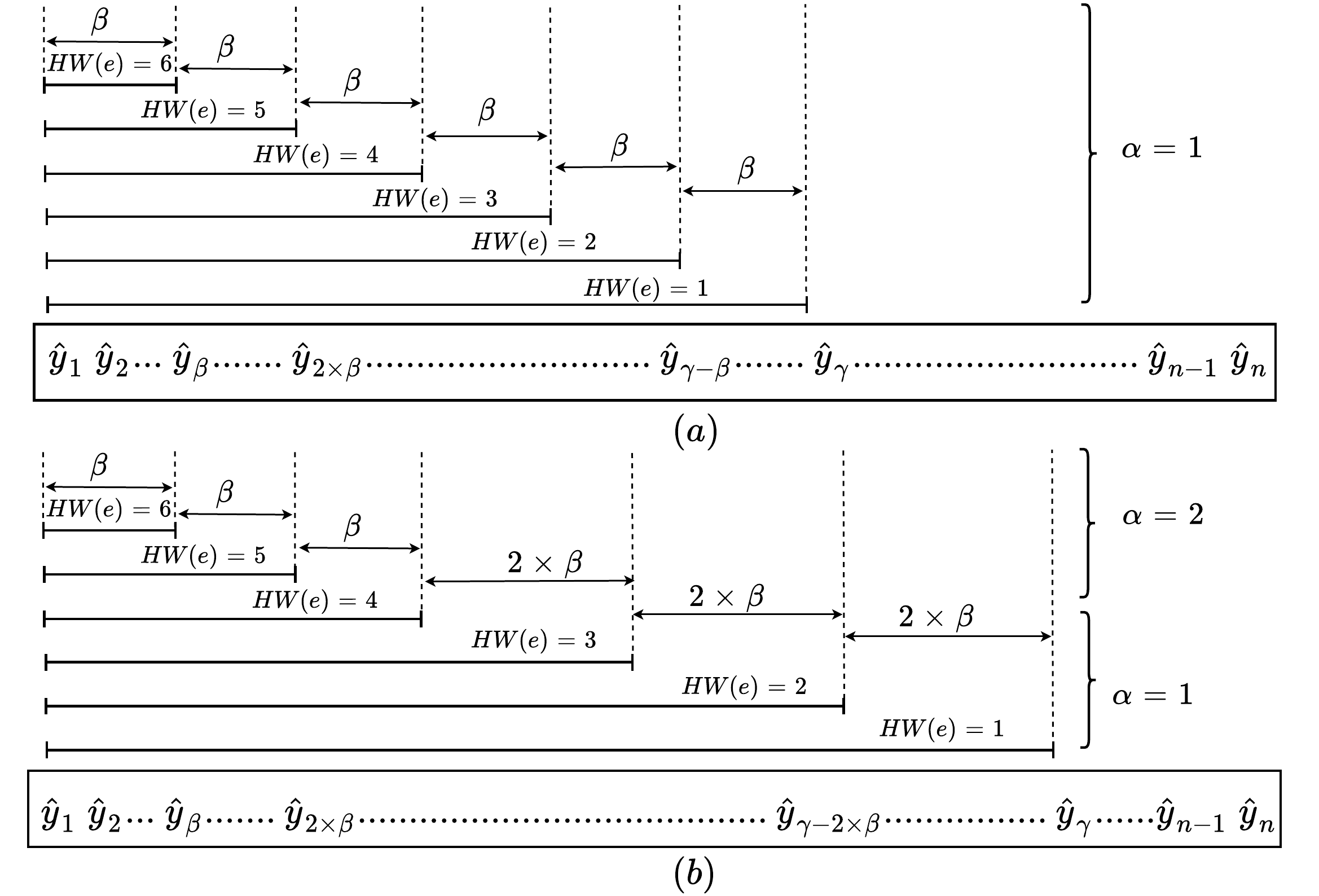}
  \vspace*{-2em}
  \caption{The subsets of $\hat{\bm{y}}$ for the proposed step-GRAND TEP generation. (a) $P = 6$ and $\alpha = 1$ (b) $P = 6$ and $\alpha = 2$}
  \label{fig:Alg_1}
  \vspace*{-1em} 
\end{figure}

\begin{algorithm}[t]
\caption{\label{alg:Stairgrand}Step-GRAND}
    \DontPrintSemicolon
    \SetAlgoVlined  
    \SetKwData{e}{$\bm{e}$}
    \SetKwData{s}{$\bm{S}$}
    \SetKwData{ind}{$\bm{ind}$}
    \SetKwData{LLR}{$\bm{y}$}
    \SetKwData{sortSet}{$[\bm{r},\bm{ind}]$}
    \SetKwData{estm}{$\hat{\bm{u}}$}
    \SetKwData{ginv}{$\bm{G}^{-1}$}
    \SetKwData{LW}{${LW_\text{max}}$}
    \SetKwData{HW}{${HW}$}
    \SetKwData{yhat}{$\hat{\bm{y}}$}
    \KwIn{\LLR, $\bm{H}$, \ginv, $P$, $\alpha$, $\beta$}
    \KwOut{\estm}
    \SetKwFunction{RecursiveComputeLLRs}{recursiveComputeLLRs}
    \SetKwFunction{HammingWeight}{HammingWeight}
    \SetKwFunction{RDecodeRONE}{redecodeR1}
    \SetKwFunction{DecodeRZERO}{decodeR0}
    \SetKwFunction{Find}{findCandidate}
    \SetKwFunction{new}{generateNewTEP}
    \SetKwFunction{intPartition}{generateAllIntPartitions}
    \SetKwFunction{Sort}{sort}
\eIf{$\bm{H} \cdot\yhat^\top == \bm{0}$}
{\KwRet{$\estm \leftarrow\yhat\cdot\ginv$}}
{
    $\ind \leftarrow$ \Sort{\LLR} \tcp*[r]{$\lvert{\bm{y}}_i\rvert\leq\lvert{\bm{y}}_j\rvert~~\forall i < j$}
    $\e \leftarrow \bm{0}$; $\HW \leftarrow 1$\;
    \For{$i \gets 1$ to $\alpha$}{
      $\gamma \leftarrow \frac{(\alpha - i + 1)\times(\alpha - i + 2)}{2}\times\frac{P}{\alpha}\times\beta$ \;
      \For{$j \gets 1$ to $\frac{P}{\alpha}$}{
        \For{$k \gets 1$ to $\binom{\gamma}{\HW}$}{
        $\e \leftarrow$ \new{\HW,$\gamma$,\ind}\;
        \If{$\bm{H} \cdot(\yhat \oplus \e)^\top == \bm{0}$} {
            $\estm \leftarrow (\yhat \oplus \e)\cdot\ginv$\;
            \KwRet{\estm}
            }
      }
      $\HW \leftarrow \HW+1$\;
      $\gamma \leftarrow \gamma-(\alpha - i +1 )\times\beta$\;
      }

    }
   } 
\end{algorithm}

\subsection{GRAND Decoding}

GRAND is centered around generating TEPs ($\bm{e}$), applying them to the hard-demodulated received vector $\bm{\hat{y}}$ and querying the resultant vector $\bm{\hat{y}}\oplus\bm{e}$ for codebook membership as follows:
\begin{equation}
\bm{H} \cdot(\bm{\hat{y}} \oplus \bm{e})^\top~=~\bm{0}
\label{eq:constraint}
\end{equation}
where $\bm{H}$ is a $(n-k)\times n$ parity check matrix of the code. If this codebook membership constraint (\ref{eq:constraint}) is satisfied, $\bm{e}$ is the guessed noise and $\hat{\bm{c}} \triangleq \bm{\hat{y}}~\oplus~\bm{e}$ is the estimated codeword. 

\section{Proposed step-GRAND } \label{sec:StairGRANDAlg} 

The proposed step-GRAND generates TEPs ($\bm{e}$) in increasing Hamming weight order, with the highest Hamming weight of the generated TEPs ($\bm{e}$) being $P$. However, unlike GRANDAB \cite{Duffy20205g}, which generates all $\binom{n}{i}$ TEPs for each Hamming weight $i~\forall~i\in [1,P]$, the step-GRAND only generates a subset of the TEPs for each Hamming weight.

The step-GRAND begins by generating $\binom{\gamma}{1}$ TEPs with a Hamming weight of 1 ($\bm{e}=\mathds{1}_i$, with $i \in \llbracket 1\isep \gamma \rrbracket$), where $\gamma$ is the size of the subset of $\hat{\bm{y}}$. After that, for the subset of $\hat{\bm{y}}$ of size $\gamma-\beta$, $\binom{\gamma-\beta}{2}$ TEPs ($\bm{e}$) with a Hamming weight of 2 ($\bm{e}=\mathds{1}_{i,j}$, with $i \in \llbracket 1\isep \gamma-\beta \rrbracket$, $j \in \llbracket 1\isep \gamma-\beta \rrbracket$ and $i\neq j$,) are generated. Please note that $\beta$, termed as \textit{step size}, refers to the size difference between two successive subsets of $\hat{\bm{y}}$. Similarly, for each subsequent Hamming weight $HW \in \llbracket 3\isep P \rrbracket$, $\binom{\gamma-(HW-1)\times\beta}{HW}$ TEPs ($\bm{e}$) are generated for the subset of $\hat{\bm{y}}$ of size $\gamma-(HW-1)\times\beta$. Figure \ref{fig:Alg_1} (a) displays the $P = 6$ subsets of $\hat{\bm{y}}$ for the proposed step-GRAND TEP generation.

The $P$ subsets of $\hat{\bm{y}}$ can also be divided into $\alpha$ segments, each of which has $\frac{P}{\alpha}$ subsets, as illustrated in Fig. \ref{fig:Alg_1} (b), where $P = 6$ and $\alpha = 2$. Furthermore, the \textit{intra-segment step size} is a multiple of $\beta$ such that the intra-segment step size of the $i^{th}$ ($i\in \llbracket 1\isep \alpha \rrbracket$) segment is $(\alpha - i + 1)\times\beta$.

The step-GRAND decoding process is summarized in Algorithm \ref{alg:Stairgrand}. The inputs to the algorithm are the vector of channel observation values $\bm{y}$ of size $n$, an $(n-k)\times n$ matrix $\bm{H}$, an $n\times k$ matrix $\bm{G}^{-1}$ such that $\bm{G}^{-1}\cdot \bm{G}$ is the $n\times n$ identity matrix, with $\bm{G}$ a generator matrix of the code, the maximum Hamming weight $P$ of TEPs ($\bm{e}$), the number of segments $\alpha$ and the step size $\beta$.

The step-GRAND sorts the received vector $\bm{y}$, Log-Likelihood Ratios (LLRs), in ascending order of absolute values of LLRs ($\lvert{\bm{y}}_i\rvert\leq\lvert{\bm{y}}_j\rvert~~\forall i < j$), and the associated indices are recorded in a permutation vector denoted by $\bm{ind}$ (line 4). The input parameters $\alpha$, $\beta$, and $P$ are used to calculate the size ($\gamma$) of a subset of $\hat{\bm{y}}$ for segment $i$ ($i\in \llbracket 1\isep \alpha \rrbracket$) (line 7). Following that, in segment $i$, all $\binom{\gamma}{HW}$ TEPs ($\bm{e}$) are generated for a particular Hamming weight $HW$ corresponding to a subset $j$ ($j\in \llbracket 1\isep \frac{P}{\alpha} \rrbracket$), with size $\gamma$ (lines 8–10). The function \textit{generateNewTEP} successively generates TEPs ($\bm{e}$), with Hamming weight $HW$, which are then ordered using the permutation vector $\bm{ind}$ (line 10). The generated TEPs ($\bm{e}$) are then applied sequentially to $\hat{\bm{y}}$ and the resulting vector ($\hat{\bm{y}}~\oplus~\bm{e}$) is then queried for codebook membership (line 11). If the codebook membership criterion (\ref{eq:constraint}) is satisfied, the original message ($\hat{\bm{u}}$) is retrieved and the decoding process is terminated (lines 12-13). Otherwise, $\gamma$, the size of the subset, is updated, and TEPs with  Hamming weight $HW+1$ are generated (lines 14-15). 


\begin{figure}[!t]
  \centering
  \begin{tikzpicture}[spy using outlines = {rectangle, magnification=2.0, connect spies}]
    \begin{groupplot}[group style={group name=fer_queries, group size= 2 by 1, horizontal sep=5pt, vertical sep=5pt},
      footnotesize,
      height=.6\columnwidth,  width=0.55\columnwidth,
      xlabel=$\frac{E_b}{N_0}$ (dB),
      xmin=0, xmax=8, xtick={0,1,...,7},
      ymode=log,
      tick align=inside,
      grid=both, grid style={gray!30},
      /pgfplots/table/ignore chars={|},
      ]

      \nextgroupplot[ylabel= FER, ytick pos=left, y label style={at={(axis description cs:-0.225,.5)},anchor=south},ymin=5e-8, ymax = 2]
      \addplot[mark=diamond , Paired-6, semithick]  table[x=Eb/N0, y=FER] {data/Polar/128_105/POLAR_N128_105_GRANDAB.txt};\label{gp:plot1_p}
      \addplot[mark=triangle, Paired-3, semithick]  table[x=Eb/N0, y=FER] {data/Polar/128_105/POLAR_N128_105_ORBGRAND.txt};\label{gp:plot2_p}
      \addplot[mark = none, Paired-7, semithick]  table[x=Eb/N0, y=FER] {data/Polar/128_105/ORBGRAND_LW64_HW6.txt};\label{gp:plot3_p}
      \addplot[mark= none , Paired-11, semithick, dashed]  table[x=Eb/N0, y=FER] {data/Polar/128_105/ORBGRAND_LW53_HW6.txt}  ;\label{gp:plot4_p}

      \addplot[mark=otimes  , Paired-5, semithick]  table[x=Eb/N0, y=FER] {data/Polar/128_105/Polar_128_105_Stair_Delta_6_updated.txt}  ;\label{gp:plot5_p}
      \addplot[mark=star  , Paired-9, semithick]  table[x=Eb/N0, y=FER] {data/Polar/128_105/Polar_128_105_Stair_Delta_6_seg_1_updated.txt}  ;\label{gp:plot6_p}
      \addplot[mark=pentagon, Paired-1, semithick]  table[x=Eb/N0, y=FER] {data/Polar/128_105/POLAR_N128_105_L_32.txt}      ;\label{gp:plot9_p};\
      \addplot[mark=+  , Paired-13 , semithick]  table[x=Eb/N0, y=FER] {data/Polar/128_105/FER_N128_105_LUT_ILW_8192.txt}; \label{gp:plot10_p}

      \coordinate (top) at (rel axis cs:0,1);

      \coordinate (spypoint1) at (axis cs:6.25,0.8e-6);
      \coordinate (magnifyglass1) at (axis cs:2.6,1.1e-5);

      \coordinate (spypoint2) at (axis cs:14.5,0.7e-6);
      \coordinate (magnifyglass2) at (axis cs:10.6,0.15e-5);

      \nextgroupplot[ylabel=Queries, ytick pos=right,y label style={at={(axis description cs:1.325,.5)},anchor=south},ymin=1, ymax = 2e7]
\addplot[mark=diamond , Paired-6, semithick]  table[x=Eb/N0, y=Tests/f] {data/Polar/128_105/POLAR_N128_105_GRANDAB.txt};
      \addplot[mark=triangle, Paired-3, semithick]  table[x=Eb/N0, y=Tests/f] {data/Polar/128_105/POLAR_N128_105_ORBGRAND.txt};
      \addplot[mark = none, Paired-7, semithick]  table[x=Eb/N0, y=Tests/f] {data/Polar/128_105/ORBGRAND_LW64_HW6.txt};
      \addplot[mark= none , Paired-11, semithick, dashed]  table[x=Eb/N0, y=Tests/f] {data/Polar/128_105/ORBGRAND_LW53_HW6.txt}  ;
      \addplot[mark=otimes  , Paired-5, semithick]  table[x=Eb/N0, y=Tests/f] {data/Polar/128_105/Polar_128_105_Stair_Delta_6_updated.txt}  ;
      \addplot[mark=star  , Paired-9, semithick]  table[x=Eb/N0, y=Tests/f] {data/Polar/128_105/Polar_128_105_Stair_Delta_6_seg_1_updated.txt}  ;
      \addplot[mark=+  , Paired-13 , semithick]  table[x=Eb/N0, y=Tests/f] {data/Polar/128_105/FER_N128_105_LUT_ILW_8192.txt};

      \coordinate (bot) at (rel axis cs:1,0);

    \end{groupplot}
    \node[below = 1cm of fer_queries c1r1.south] {\footnotesize (a) : FER};
    \node[below = 1cm of fer_queries c2r1.south] {\footnotesize (b) : Avg. Queries};
    \path (top|-current bounding box.north) -- coordinate(legendpos) (bot|-current bounding box.north);
    \matrix[
    matrix of nodes,
    anchor=south,
    draw,
    inner sep=0.2em,
    draw
    ]at(legendpos)
    {
    \ref{gp:plot6_p}& \tiny Step-GRAND ($\alpha=1$, $\beta=6$, $P=6$) &[1pt]
      \ref{gp:plot1_p}& \tiny GRANDAB ($AB=3$) \\
      \ref{gp:plot5_p}& \tiny Step-GRAND ($\alpha=2$, $\beta=6$, $P=6$) &[1pt]
      \ref{gp:plot9_p}& \tiny SCL ($L = 32$) \\
      \ref{gp:plot3_p}& \tiny ORBGRAND ($LW_\text{max}=64$, $P=6$) &[1pt] 
      \ref{gp:plot2_p}& \tiny ORBGRAND ($LW_\text{max}=8256$) \\
      \ref{gp:plot4_p}& \tiny ORBGRAND ($LW_\text{max}=53$, $P=6$) &[1pt] 
      \ref{gp:plot10_p}& \tiny F.L ORBGRAND \cite{carloORB} (Queries$_\text{max}=2^{13}$) \\ 
      }; 
  \end{tikzpicture}
  \vspace*{-2em}
  \caption{\label{fig:FER_polar} Comparison of decoding performance and average complexity of different GRAND variants for polar code $(128,105+11)$.}
   \vspace*{-1.1em}
\end{figure}
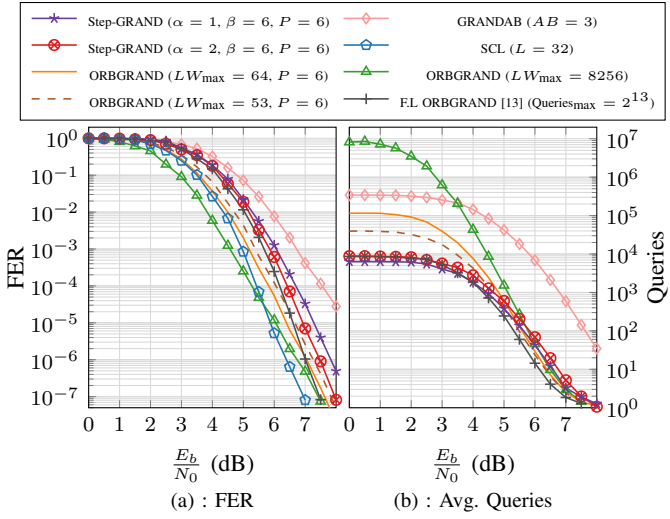


\begin{figure}[!t]
  \centering
  \begin{tikzpicture}[spy using outlines = {rectangle, magnification=2.0, connect spies}]
    \begin{groupplot}[group style={group name=fer_queries, group size= 2 by 1, horizontal sep=5pt, vertical sep=5pt},
      footnotesize,
      height=.6\columnwidth,  width=0.55\columnwidth,
      xlabel=$\frac{E_b}{N_0}$ (dB),
      xmin=0, xmax=8, xtick={0,1,...,7},
      ymode=log,
      tick align=inside,
      grid=both, grid style={gray!30},
      /pgfplots/table/ignore chars={|},
      ]

      \nextgroupplot[ylabel= FER, ytick pos=left, y label style={at={(axis description cs:-0.225,.5)},anchor=south},ymin=5e-8, ymax = 2]
      \addplot[mark=diamond , Paired-6, semithick]  table[x=Eb/N0, y=FER] {data/BCH/127_106/BCH_N127_106_BM.txt};\label{gp:plot1_b}
      \addplot[mark=triangle, Paired-3, semithick]  table[x=Eb/N0, y=FER] {data/BCH/127_106/BCH_N127_106_ORBGRAND.txt};\label{gp:plot2_b}
      \addplot[mark = none, Paired-7, semithick]  table[x=Eb/N0, y=FER] {data/BCH/127_106/ORBGRAND_LW64_HW6.txt};\label{gp:plot3_b}
      \addplot[mark= none , Paired-11, semithick, dashed]  table[x=Eb/N0, y=FER] {data/BCH/127_106/ORBGRAND_LW56_HW6.txt}  ;\label{gp:plot4_b}

      \addplot[mark=otimes  , Paired-5, semithick]  table[x=Eb/N0, y=FER] {data/BCH/127_106/BCH_127_106_Stair_Delta_7_updated.txt}  ;\label{gp:plot5_b}
      \addplot[mark=star  , Paired-9, semithick]  table[x=Eb/N0, y=FER] {data/BCH/127_106/BCH_127_106_Stair_Delta_7_seg_1_updated.txt}  ;\label{gp:plot6_b}
      \addplot[mark=pentagon, Paired-1, semithick]  table[x=Eb/N0, y=FER] {data/BCH/127_106/BCH_N127_106_ML.txt}      ;\label{gp:plot9_b};\
      \addplot[mark=+  , Paired-13 , semithick]  table[x=Eb/N0, y=FER] {data/BCH/127_106/FER_N127_106_ILW_10_4.txt}; \label{gp:plot10_b}

      \coordinate (top) at (rel axis cs:0,1);

      \coordinate (spypoint1) at (axis cs:6.25,0.8e-6);
      \coordinate (magnifyglass1) at (axis cs:2.6,1.1e-5);

      \coordinate (spypoint2) at (axis cs:14.5,0.7e-6);
      \coordinate (magnifyglass2) at (axis cs:10.6,0.15e-5);

      \nextgroupplot[ylabel=Queries, ytick pos=right,y label style={at={(axis description cs:1.325,.5)},anchor=south},ymin=1, ymax = 5e6]
      \addplot[mark=triangle, Paired-3, semithick]  table[x=Eb/N0, y=Tests/f] {data/BCH/127_106/BCH_N127_106_ORBGRAND.txt};
      \addplot[mark = none, Paired-7, semithick]  table[x=Eb/N0, y=Tests/f] {data/BCH/127_106/ORBGRAND_LW64_HW6.txt};
      \addplot[mark= none , Paired-11, semithick, dashed]  table[x=Eb/N0, y=Tests/f] {data/BCH/127_106/ORBGRAND_LW56_HW6.txt}  ;
      \addplot[mark=otimes  , Paired-5, semithick]  table[x=Eb/N0, y=Tests/f] {data/BCH/127_106/BCH_127_106_Stair_Delta_7_updated.txt}  ;
      \addplot[mark=star  , Paired-9, semithick]  table[x=Eb/N0, y=Tests/f] {data/BCH/127_106/BCH_127_106_Stair_Delta_7_seg_1_updated.txt}  ;
      \addplot[mark=+  , Paired-13 , semithick]  table[x=Eb/N0, y=Tests/f] {data/BCH/127_106/FER_N127_106_ILW_10_4.txt};

      \coordinate (bot) at (rel axis cs:1,0);

    \end{groupplot}
    \node[below = 1cm of fer_queries c1r1.south] {\footnotesize (a) : FER};
    \node[below = 1cm of fer_queries c2r1.south] {\footnotesize (b) : Avg. Queries};
    \path (top|-current bounding box.north) -- coordinate(legendpos) (bot|-current bounding box.north);
    \matrix[
    matrix of nodes,
    anchor=south,
    draw,
    inner sep=0.2em,
    draw
    ]at(legendpos)
    {
    \ref{gp:plot6_b}& \tiny Step-GRAND ($\alpha=1$, $\beta=7$, $P=6$) &[1pt]
      \ref{gp:plot1_b}& \tiny B-M Decoder \\
      \ref{gp:plot5_b}& \tiny Step-GRAND ($\alpha=2$, $\beta=7$, $P=6$) &[1pt]
      \ref{gp:plot9_b}& \tiny ML \cite{kaiserslautern} \\
      \ref{gp:plot3_b}& \tiny ORBGRAND ($LW_\text{max}=64$, $P=6$) &[1pt] 
      \ref{gp:plot2_b}& \tiny ORBGRAND ($LW_\text{max}=8128$)  \\
      \ref{gp:plot4_b}& \tiny ORBGRAND ($LW_\text{max}=56$, $P=6$) &[1pt]
      \ref{gp:plot10_b}& \tiny ORBGRAND-ILWO \cite{carloILW} (Queries$_\text{max}=10^{4}$)\\ 
      }; 
  \end{tikzpicture}
  \vspace*{-2em}
  \caption{\label{fig:FER_bch} Comparison of decoding performance and average complexity of different GRAND variants for BCH $(127,106)$ code.}
   \vspace*{-1.3em}
\end{figure}
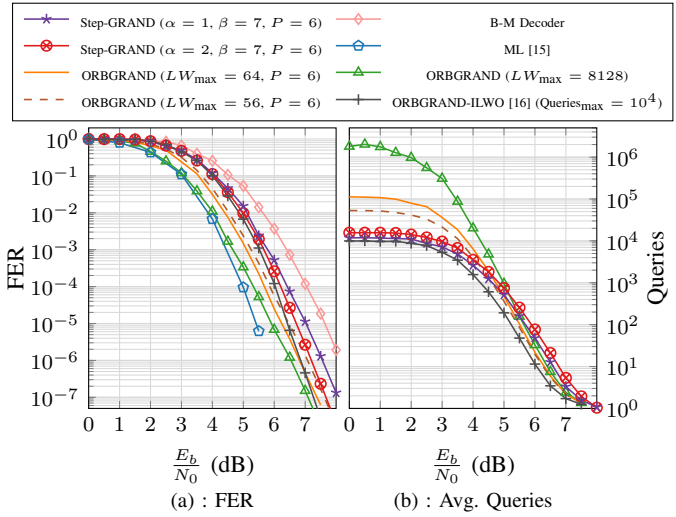

\section{Performance Evaluation}\label{sec:performance} 

Figures \ref{fig:FER_polar} and \ref{fig:FER_bch} provide a comparison of the decoding performance as well as the computational complexity, expressed as the number of codebook membership queries required, of step-GRAND with other GRAND variants for the respective 5G NR CRC-Aided (CA)-polar code $(128,105+11)$ and Bose-Chaudhuri-Hocquenghem (BCH) code $(127,106)$ \cite{Hocquenghem59,Bose1960}. Furthermore, the decoding performance of the Improved Logistics Weight Order (ILWO) ORBGRAND decoder \cite{carloORB}, CA-Successive Cancellation List (CA-SCL) decoder \cite{Tal15} and ML decoder \cite{kaiserslautern} are presented for reference. Please note that, a BPSK modulation over an AWGN channel with variance $\sigma^2$ is considered for the numerical simulation results presented in this section. As demonstrated in Fig. \ref{fig:FER_polar} and \ref{fig:FER_bch}, the proposed step-GRAND decoder outperforms the hard-input GRANDAB decoder \cite{Duffy19TIT} and the Berlekamp-Massey (B-M) \cite{Berlekamp68}\cite{Massey69} decoder. Furthermore, the decoding performance of step-GRAND approaches that of ORBGRAND \cite{duffy2020ordered} with improved channel conditions and various parametric settings.

The parameters maximum logistic weight ($LW_\text{max}$) and $P$ impact both the decoding performance and computational complexity of the baseline ORBGRAND \cite{duffy2020ordered}\cite{ORBGRAND-TVLSI}. However, the decoding performance and computational complexity of the proposed step-GRAND are influenced by the parameters $\alpha$, $\beta$ and $P$ (Alg. \ref{alg:Stairgrand}). The worst-case complexity for the proposed step-GRAND is $\sum\limits_{HW=1}^{P}\binom{\gamma}{HW}$, where $\gamma$ is the size of the subset of $\hat{\bm{y}}$ for which TEPs with a Hamming weight $HW~(HW \in [1, P])$ are evaluated.

At a target FER of $\leq10^{-7}$, the proposed step-GRAND ($\alpha=2$, $\beta=6$, $P=6$) achieves similar decoding performance to ORBGRAND ($LW_\text{max}=53$, $P=6$), but experiences a degradation in decoding performance of $0.3$ dB when compared to ORBGRAND ($LW_\text{max}=64$, $P=6$), as shown in Fig. \ref{fig:FER_polar} (a). The Worst-Case (W.C) complexity for ORBGRAND ($LW_\text{max}=53$, $P=6$) is $3.92\times10^{4}$ queries \cite{ListGRAND-TVLSI}, whereas the W.C complexity for step-GRAND ($\alpha=2$, $\beta=6$, $P=6$) is \numprint{8828} queries\footnote{ With parameters ($\alpha=2$, $\beta=6$, $P=6$), the $(\gamma,HW)$ values are $(54,1)$,$(42,2)$,$(30,3)$,$(18,4)$,$(12,5)$ and $(6,6)$. (Sec. \ref{sec:StairGRANDAlg})}. Therefore, the W.C complexity of the proposed step-GRAND is $\frac{1}{4}\times$ that of ORBGRAND ($LW_\text{max}=53$, $P=6$). Similarly, when decoding BCH code $(127,106)$ at a target FER of $\leq10^{-7}$, the step-GRAND ($\alpha=2$, $\beta=7$, $P=6$) achieves similar decoding performance to ORBGRAND ($LW_\text{max}=56$, $P=6$) but experiences a performance degradation of $0.2$ dB when compared to ORBGRAND ($LW_\text{max}=64$, $P=6$), as shown in Fig. \ref{fig:FER_bch} (a). However, step-GRAND ($\alpha=2$, $\beta=7$, $P=6$) has a W.C complexity of \numprint{15778} queries while ORBGRAND ($LW_\text{max}=56$, $P=6$) has a W.C complexity of $5.37\times10^{4}$ queries \cite{ListGRAND-TVLSI}; as a result, the W.C complexity of step-GRAND ($\alpha=2$, $\beta=7$, $P=6$) is $\frac{1}{3}\times$ that of ORBGRAND ($LW_\text{max}=56$, $P=6$).

To summarize, the parameters of the proposed step-GRAND ($\alpha$, $\beta$, $P$) can be adjusted for various classes of channel codes in order to achieve a balance between decoding performance requirements and the complexity/latency budget for a target application.

\begin{figure}[!tb]
  \centering
  \includegraphics[width=1\linewidth]{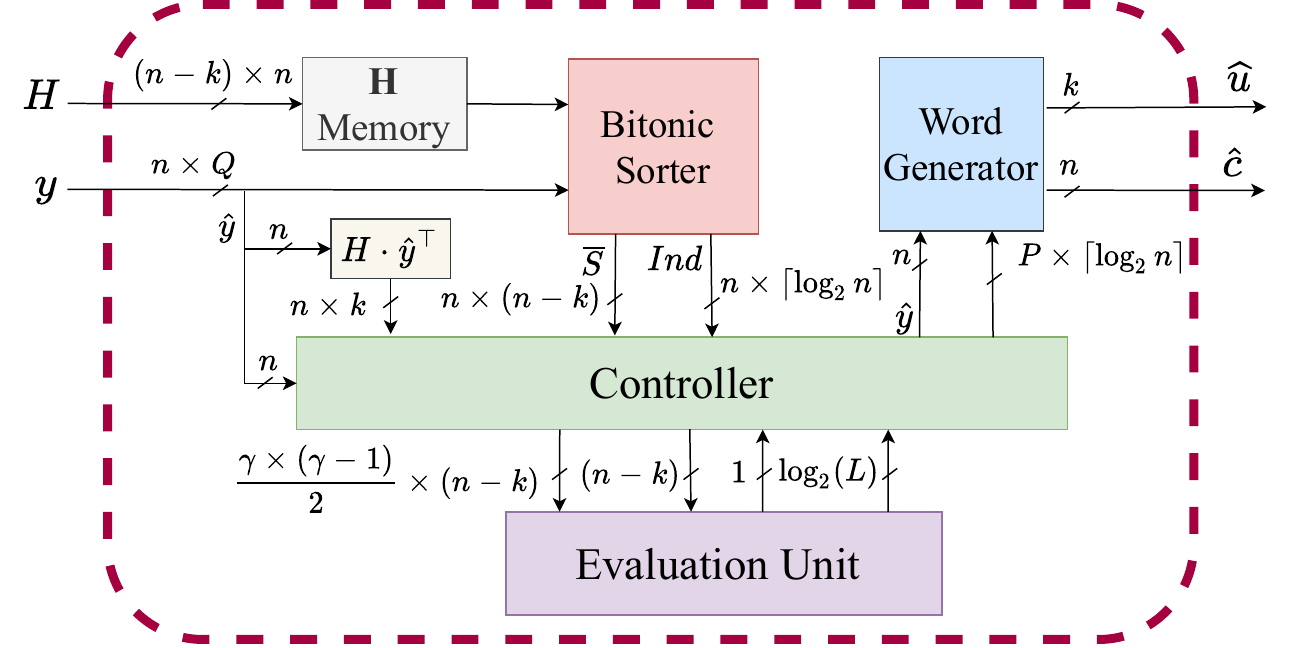}
  \vspace*{-2em}
  \caption{Proposed VLSI architecture for step-GRAND}
  \label{fig:StairGRAND_VLSI} 
  \vspace*{-1em} 
\end{figure}
\begin{figure}[!tb]
  \centering
  \includegraphics[width=1\linewidth]{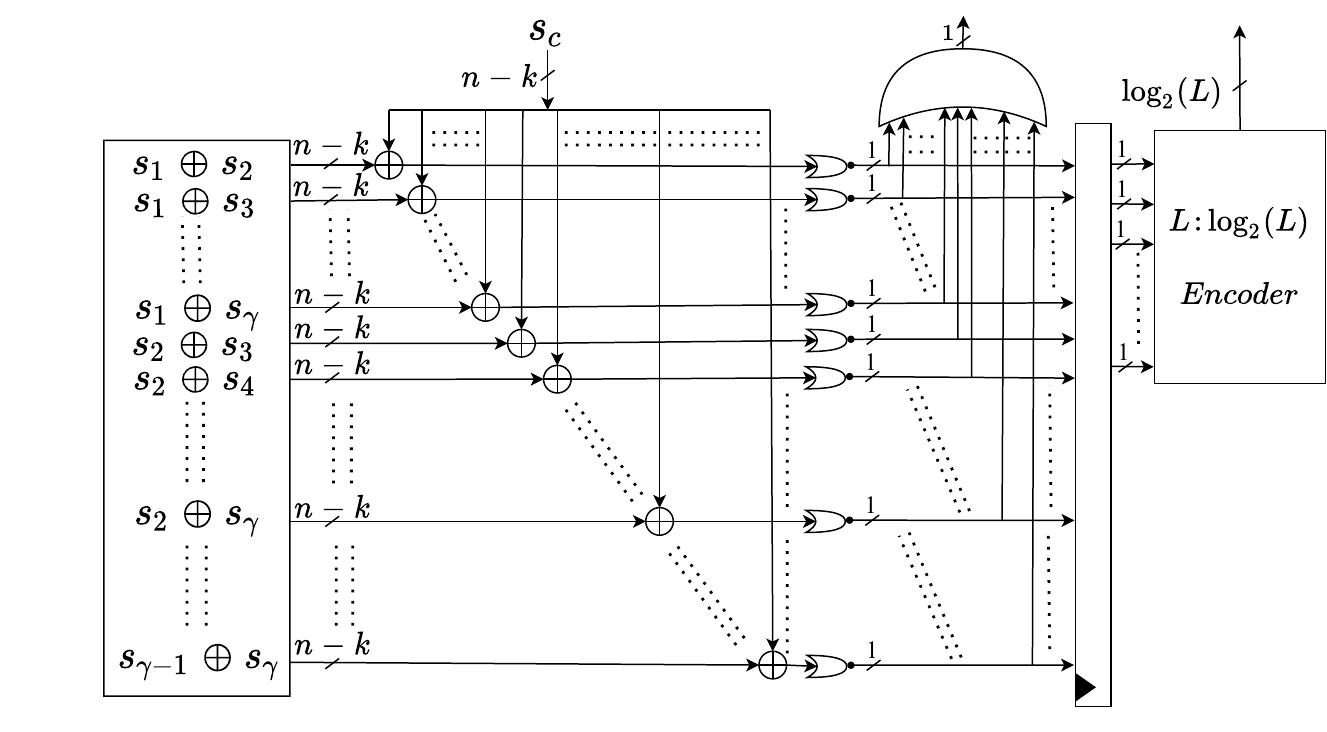}
  \vspace*{-2em}
  \caption{Proposed \textit{evaluation unit} for step-GRAND}
  \label{fig:2bf_vlsi}
  \vspace*{-1em} 
\end{figure}
\begin{figure}[!tb]
  \centering
  \includegraphics[width=0.95\linewidth]{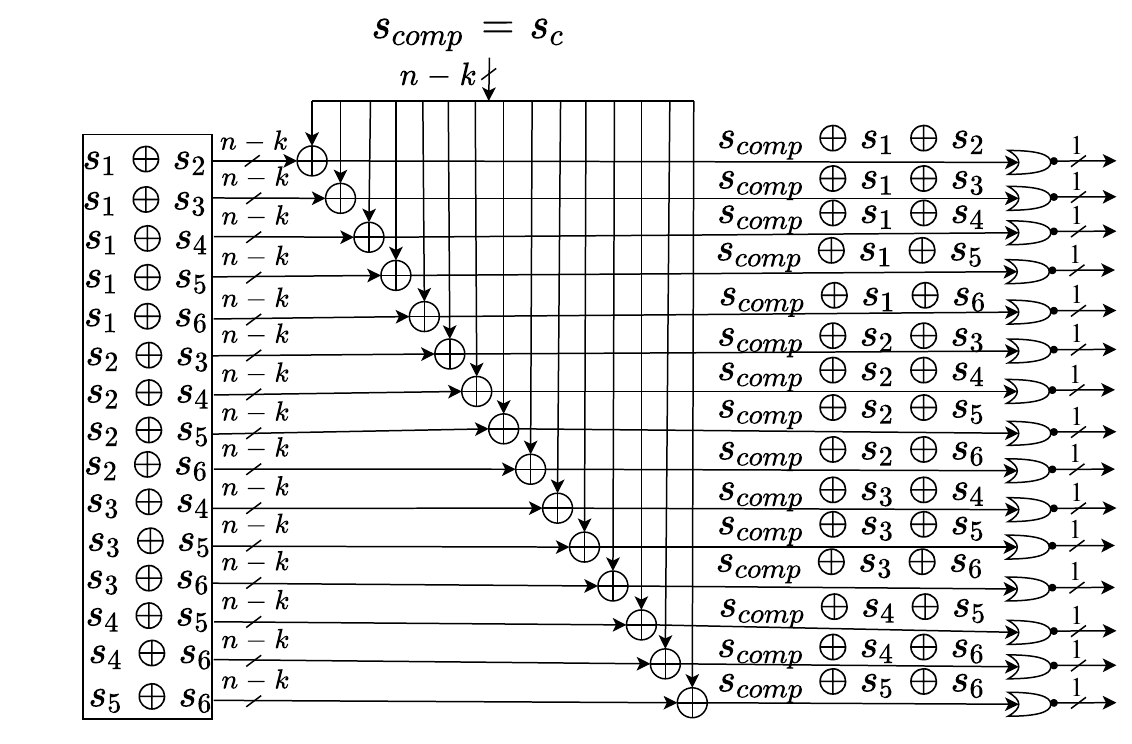}
  \vspace*{-1.25em}
  \caption{Evaluating TEPs with a Hamming weight of 2 for $\gamma = 6$.}
  \label{fig:2bf_vlsi_micro}
  \vspace*{-1em} 
\end{figure}
\begin{figure}[!tb]
  \centering
  \includegraphics[width=0.8\linewidth]{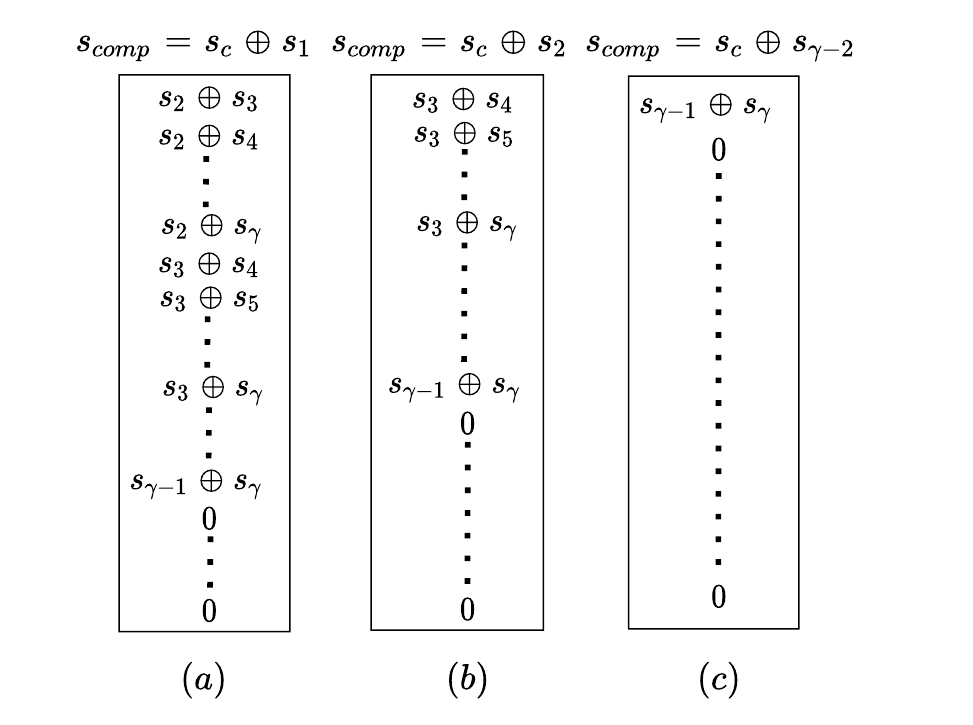}
  \vspace*{-1em}
  \caption{Evaluating TEPs with a Hamming weight of 3}
  \label{fig:3bf_sch}
  \vspace*{-1em} 
\end{figure}
\begin{figure}[!tb]
  \centering
  \includegraphics[width=0.95\linewidth]{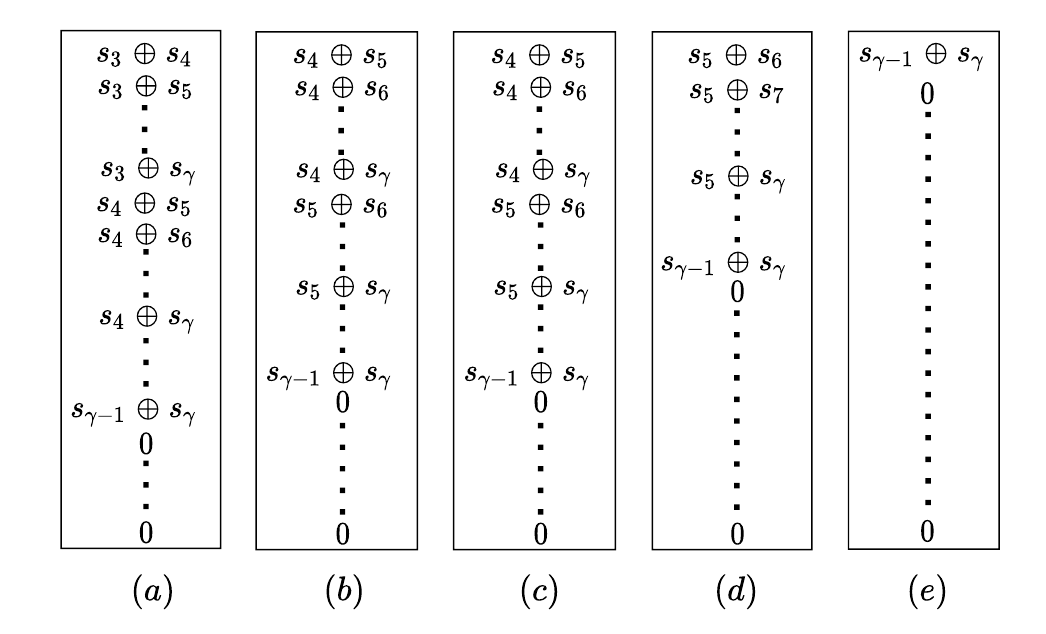}
  \vspace*{-1em}
  \caption{Evaluating TEPs with a Hamming weight of 4 (a) $\bm{s_{comp} = \bm{s_c}\oplus\bm{s}_1\oplus\bm{s}_2}$ (b) $\bm{s_{comp} = \bm{s_c}\oplus\bm{s}_1\oplus\bm{s}_3}$ (c) $\bm{s_{comp} = \bm{s_c}\oplus\bm{s}_2\oplus\bm{s}_3}$ (d) $\bm{s_{comp} = \bm{s_c}\oplus\bm{s}_3\oplus\bm{s}_4}$ (e) $\bm{s_{comp}} =\bm{s_c}\oplus\bm{s}_{\gamma-3}\bm{s}_{\gamma-2}$}
  \label{fig:4bf_sch}
  \vspace*{-1em} 
\end{figure}

\section{VLSI Architecture for step-GRAND} \label{sec:StairGRANDVLSI}

In this section, we present the VLSI architecture for step-GRAND, which is designed for universal decoding of $(n,k)$ linear block codes. The proposed hardware architecture builds upon the previously proposed hard-input GRANDAB hardware \cite{GRANDAB-VLSI} and soft-input ORBGRAND hardware \cite{ORBGRAND-TVLSI}, both of which use shift registers to store the $(n-k)$-bit syndromes associated with TEPs with a Hamming weight of 1 (denoted as $\bm{s}_i=\bm{H}\cdot\mathds{1}_i^\top,~i\in \llbracket 1\isep n \rrbracket$). Furthermore, we leverage the linearity property of the underlying code to combine $l$ TEP syndromes, corresponding to error patterns with Hamming weight of $1$ ($\bm{s}_i$), to generate syndromes corresponding to an error pattern with a Hamming weight of $l$ ($\bm{s}_{1,2\ldots,l} = \bm{H}\cdot\mathds{1}_1^\top \oplus \bm{H}\cdot\mathds{1}_2^\top \ldots \oplus\bm{H}\cdot\mathds{1}_l^\top $). We refer the reader to \cite{GRANDAB-VLSI} and \cite{ORBGRAND-TVLSI} for more details.

The proposed VLSI architecture for the step-GRAND, which can be used to decode any linear block code with a length of $n$ and a coding rate of $0.75\leq R \leq 1$, is shown in Fig.~\ref{fig:StairGRAND_VLSI}. Any parity check matrix $(H)$ can be loaded into $(n-k)\times n\text{-bit}$ \textit{H memory} to support various classes of channel codes. The proposed step-GRAND hardware receives soft channel observations values (LLRs) ${\bm{y}}$ from the communication channel as an input and then applies the codebook membership verification (\ref{eq:constraint}) to the hard-decided vector $\hat{\bm{y}}$. The decoding is terminated if the codebook membership criterion (\ref{eq:constraint}) is satisfied for $\hat{\bm{y}}$ ($\bm{s_c}=\bm{H}\cdot\bm{\hat{y}}^\top = \bm{0}$). Otherwise, the \textit{bitonic sorter} \cite{batcher68} is employed to sort the LLRs (${\bm{y}}$) in ascending order of their absolute values ($\lvert{\bm{y}}_i\rvert\leq\lvert{\bm{y}}_j\rvert~~\forall i < j$). The bitonic sorter is pipelined to $\log_2(n)$ stages, and thus, sorting the received LLRs (${\bm{y}}$) takes $\log_2(n)$ clock cycles. Following that, in a single time step, the codebook membership of all TEPs ($\bm{e}=\mathds{1}_i$) with a Hamming weight of 1 is evaluated ($\bm{s_c}\oplus\bm{s}_i = \bm{0}, \forall~i\in [1, \gamma]$).

The TEPs with Hamming weight $HW>1$ ($\forall ~HW \in [2, P]$) are evaluated for codebook membership by the \textit{controller} module in conjunction with the \textit{evaluation unit}. If any of the evaluated TEPs ($\bm{e}=\mathds{1}_{1,2\ldots,HW}$) satisfy the codebook membership constraint ($\bm{s_c}\oplus\bm{s}_{1,2\ldots,HW}=\bm{0}$), a 2D \textit{priority encoder} module is used in conjunction with the controller module to pass the corresponding indices to the \textit{word generator} module, which maps the sorted index values ($\bm{ind}$) to the appropriate bit flip locations in $\hat{\bm{c}}$.

\subsection{Scheduling and Details}
Figure \ref{fig:2bf_vlsi} depicts the microarchitecture of the \textit{evaluation unit} of the proposed step-GRAND, which employs a $\binom{\gamma}{2}\times(n-k)$-bit shift register to store $\binom{\gamma}{2}$ syndromes of TEPs with a Hamming weight of 2 ($\bm{s}_{i,j},\forall~i\in \llbracket 1\isep \gamma-1 \rrbracket, ~j\in \llbracket i+1\isep \gamma \rrbracket$). The generated test syndromes ($\bm{s_c}\oplus\bm{s}_{i,j}$) are NOR reduced to evaluate all the $\binom{\gamma}{2}$ TEPs for codebook membership in parallel, as shown in Fig. \ref{fig:2bf_vlsi_micro} for $\gamma = 6$ and $HW = 2$. With the proposed evaluation unit, it only requires one time-step to evaluate, for codebook membership, all $\binom{\gamma}{2}$ TEPs with a Hamming weight of 2. The $L$-to-$\log_2L$ priority encoder (shown in Fig. \ref{fig:2bf_vlsi})  is enabled to output the corresponding indices to the controller module, in $\frac{\gamma\times(\gamma-1)}{2\times L}$ time-steps, if and only if the tested syndromes satisfy the codebook membership criteria ($\bm{s_c}\oplus\bm{s}_{i,j} = \bm{0}$).

To evaluate all TEPs corresponding to Hamming weight $3 \leq HW \leq P$, the controller module generates the composite syndrome $\bm{s_{comp} = \bm{s_c}\oplus\bm{s}_{1,2\ldots,P-2}}$, which is combined with the syndromes stored in the shift register. Figure \ref{fig:3bf_sch}(a) displays the contents of the shift register and the composite syndrome $\bm{s_{comp} = \bm{s_c}\oplus\bm{s}_1}$ generated by the controller to evaluate the first set of $\frac{(\gamma-1)\times(\gamma-2)}{2}$ TEPs with a Hamming weight of 3. The shift register is shifted-up by $\gamma-2$ at the next time step, and the controller generates $\bm{s_{comp} = \bm{s_c}\oplus\bm{s}_2}$ to evaluate the subsequent $\frac{(\gamma-2)\times(\gamma-3)}{2}$ TEPs as shown in Fig. \ref{fig:3bf_sch}(b). This procedure is repeated until the controller generates $\bm{s_{comp} = \bm{s_c}\oplus\bm{s}_{\gamma-2}}$ and the final TEP with Hamming weight of 3 ($\bm{s_c}\oplus\bm{s}_{\gamma-2}\oplus\bm{s}_{\gamma-1}\oplus\bm{s}_{\gamma}$) is evaluated, as illustrated in Fig. \ref{fig:3bf_sch} (c). Therefore, evaluating all $\binom{\gamma}{3}$  TEPs with a Hamming weight of 3 requires $\binom{\gamma-2}{1}$ time steps, where the controller outputs $\bm{s_{comp} = \bm{s_c}\oplus\bm{s}_i},\forall~i\in \llbracket 1\isep \gamma-2 \rrbracket$ at each time step and the shift register is shifted up by $\gamma-i-1,\forall~i\in \llbracket 1\isep \gamma-2 \rrbracket$.

To evaluate the TEPs with a Hamming weight of 4, the controller generates $\bm{s_{comp} = \bm{s_c}\oplus\bm{s}_1\oplus\bm{s}_2}$ in the following time step. Figure \ref{fig:4bf_sch} (a) shows the content of the shift register required to evaluate $\frac{(\gamma-2)\times(\gamma-3)}{2}$ TEPs with $\bm{s_{comp} = \bm{s_c}\oplus\bm{s}_1\oplus\bm{s}_2}$ generated by the controller. The shift register is shifted up by $\gamma-3$ in the subsequent time step, as shown in Fig. \ref{fig:4bf_sch} (b), and the controller generates $\bm{s_{comp} = \bm{s_c}\oplus\bm{s}_1\oplus\bm{s}_3}$ to evaluate the next $\frac{(\gamma-3)\times(\gamma-4)}{2}$ TEPs. In $\gamma-3$ time steps, all TEPs with a Hamming weight of 4 and $\bm{s_{comp} = \bm{s_c}\oplus\bm{s}_1\oplus\bm{s}_i},\forall~i\in \llbracket 2\isep \gamma-3 \rrbracket$ are evaluated.  

The controller generates $\bm{s_{comp} = \bm{s_c}\oplus\bm{s}_2\oplus\bm{s}_3}$ in the next time step, and the contents of the shift registers are shown in Fig. \ref{fig:4bf_sch} (c). This configuration evaluates the set of $\frac{(\gamma-4)\times(\gamma-5)}{2}$ TEPs, with a Hamming weight of 4 and $\bm{s_{comp} = \bm{s_c}\oplus\bm{s}_2\oplus\bm{s}_3}$ output by the controller. The shift register is shifted up $\gamma-4$ in the following time step and controller generates $\bm{s_{comp} = \bm{s_c}\oplus\bm{s}_2\oplus\bm{s}_3}$ to evaluate the next $\frac{(\gamma-5)\times(\gamma-6)}{2}$ TEPs as shown in Fig. \ref{fig:4bf_sch} (d). This process is repeated, and after $\gamma-4$ time-steps, all TEPs with $\bm{s_{comp} = \bm{s_c}\oplus\bm{s}_2\oplus\bm{s}_i},\forall~i\in \llbracket 3\isep \gamma-4 \rrbracket$ are evaluated. This procedure continues, and the controller module generates $\bm{s_{comp}} =\bm{s_c}\oplus\bm{s}_{\gamma-3}\bm{s}_{\gamma-2}$ as shown in Fig. \ref{fig:4bf_sch} (e) in order to evaluate the final TEP with a Hamming weight of 4. We can infer from the prior explanation that it takes $\binom{\gamma-2}{2}$ time steps to evaluate all $\binom{\gamma}{4}$ TEPs with a Hamming weight of 4. 

The preceding discussion can be summarized as, by leveraging the proposed hardware, TEPs with Hamming weights $3 \leq HW \leq P$ can be evaluated in $\sum\limits_{HW=3}^{P}\binom{\gamma-2}{HW-2}$ time steps. Hence, the worst-case latency of the proposed step-GRAND hardware is given by:
\begin{equation}
\label{eq:nb_steps}
3 + \log_2(n) + \sum\limits_{HW=3}^{P}\binom{\gamma-2}{HW-2}.
\end{equation}
where $\log_2(n)$ is the latency of the sorter module.

\begin{table}[!t] 
\centering
\caption{\label{table:StairGRANDPolar}TSMC 65 nm CMOS Synthesis Comparison results for step-GRAND.}
\vspace*{-1em}
\begin{adjustbox}{max width=\columnwidth}
\begin{tabular}{lcccc}
\toprule
  & This Work  & \cite{ORBGRAND-TVLSI}  &  \cite{carloORB} &  \cite{ORBBU} \\
  \cmidrule(l){2-2}\cmidrule(l){3-3}\cmidrule(l){4-4}\cmidrule(l){5-5}
\multirow{4}{*}{Parameters} & $\alpha\leq2$ & LW$\leq$64  & ${Q_\text{max}}  = 2^{13}$ & LW$\leq$104  \\  
            & $\beta\leq6$ &  & $Q_\text{LUT} = 512$ &  \\       
            & $P\leq6$ & $P\leq6$ & $Q_\text{S} = 256$ & P$\leq$13 \\
            &  &  & $T=34$  &  \\
Implementation &      \multicolumn{3}{c} {Synthesis}     &    Fabricated     \\
Technology (nm) & 65 & 65 & 7 & 40 \\
Supply (V) & 0.9 & 0.9 & 0.5 & 1.0 \\
Quantization (bits) & 5 & 5 & N.R & 6 \\
Code length $(n)$ & 128 & 128 & 128 & 256 \\
Max. Frequency (MHz) & 454 & 454 & 701 & 90 \\
Area (mm\textsuperscript{2}) & 1.18 & 1.82 & 3.70 & 0.4 \\
W.C. Latency (ns) & 614.5 & 4224.6$^a$ & 58.49 & N.R \\
Avg. Latency (ns) & 2.2$^b$ & 2.2$^b$ & 58.49 & 40$^c$ \\
W.C. T/P (Mbps) & 170.8$^e$ & 24.85$^e$ & 73610$^e$ & N.R$^d$ \\
Avg. T/P (Gbps) & 47.7$^e$ & 47.7$^e$ & 73.61$^e$ & 6.5$^f$ \\
Code compatible & Yes & Yes & Yes & Yes \\
Rate compatible & Yes & Yes & Yes & Yes \\
\bottomrule
\multicolumn{5}{l}{\footnotesize $^a$ Corresponding to parameters $LW=53$ and $P=6$ (Fig. \ref{fig:FER_polar} (a)) }\\
\multicolumn{5}{l}{\footnotesize $^b$ For $\frac{E_b}{N_0}$ $\geq8$dB (Fig. \ref{fig:lat_tp}), $^c$ At the target FER of $10^{-7}$, $^d$N.R: Not Reported}\\
\multicolumn{5}{l}{\footnotesize $^e$ For CA-Polar Code (128,105), $^f$ For CA-Polar Code (256,240).} \\
\end{tabular}
\end{adjustbox}
\vspace*{-1em}
\end{table}


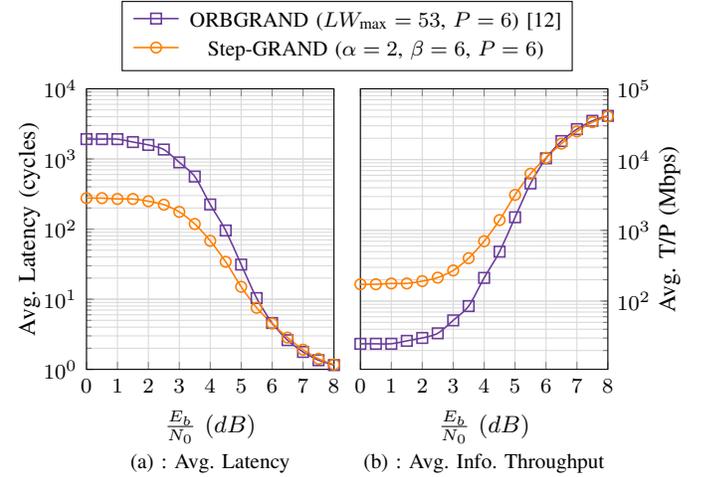
\begin{figure}[!t]
\centering
  \begin{tikzpicture}
    \begin{groupplot}[group style={group name=lat_tp, group size= 2 by 1, horizontal sep=10pt, vertical sep=10pt}, 
                      footnotesize,
                      height=.6\columnwidth,  width=.55\columnwidth,
                      xlabel=$\frac{E_{b}}{N_{0}}$ $(dB)$,
                      xmin=0, xmax=8, xtick={0,1,...,8},
                      ymode=log,
                      tick align=inside, 
                      grid=both, grid style={gray!30},
             ]

      \nextgroupplot[ylabel= Avg. Latency (cycles), ytick pos=left, y label style={at={(axis description cs:-0.15,.5)},anchor=south}, ymin=1, ymax = 1e4]
        \addplot[mark=square, smooth, Paired-9, semithick]  table[x=SNR, y=LAT] {data/Polar/128_105/Lat_ORB_128_105_LW53.txt};\label{gp:plot1_lat}
        \addplot[mark=o, smooth,Paired-7, semithick]  table[x=SNR, y=LAT] {data/Polar/128_105/Lat_Step_128_105.txt};\label{gp:plot2_lat}

        \coordinate (top) at (rel axis cs:0,1);

      \nextgroupplot[ylabel=Avg. T/P (Mbps), ytick pos=right,y label style={at={(axis description cs:1.33,.5)},anchor=south}, ,ymax = 1e5]
        \addplot[mark=square, smooth, Paired-9, semithick]  table[x=SNR, y=TP] {data/Polar/128_105/Lat_ORB_128_105_LW53.txt};
        \addplot[mark=o, smooth, Paired-7, semithick]  table[x=SNR, y=TP] {data/Polar/128_105/Lat_Step_128_105.txt};

        \coordinate (bot) at (rel axis cs:1,0);
    \end{groupplot}
    \node[below = 1cm of lat_tp c1r1.south] {\footnotesize (a) : Avg. Latency};
    \node[below = 1cm of lat_tp c2r1.south] {\footnotesize (b) : Avg. Info. Throughput};
    \path (top|-current bounding box.north) -- coordinate(legendpos) (bot|-current bounding box.north);
    \matrix[
        matrix of nodes,
        anchor=south,
        draw,
        inner sep=0.2em,
        draw
      ]at(legendpos)
      {
        \ref{gp:plot1_lat}& \footnotesize ORBGRAND ($LW_\text{max}=53$, $P=6$) \cite{ORBGRAND-TVLSI}  \\
        \ref{gp:plot2_lat}& \footnotesize Step-GRAND ($\alpha=2$, $\beta=6$, $P=6$) \\
        };
  \end{tikzpicture}
  \vspace*{-2em}
  \caption{\label{fig:lat_tp}Comparison of average latency and average information throughput for the ORBGRAND\cite{ORBGRAND-TVLSI} VLSI architecture and the proposed step-GRAND architecture for Polar code (128,105+11).}
  \vspace*{-1em}
\end{figure}

\subsection{Implementation Results}

The proposed step-GRAND VLSI architecture, with parameters ($\alpha=2$, $\beta=6$, $P=6$), has been implemented in Verilog HDL and synthesized with Synopsys Design Compiler using general-purpose TSMC 65 nm CMOS technology. Table \ref{table:StairGRANDPolar} presents the synthesis results for the proposed step-GRAND implementation, with $n=128$, and code-rate $0.75\leq~R~\leq1$, and compares them with state-of-the-art ORBGRAND hardware implementations\cite{ORBGRAND-TVLSI}\cite{carloORB}\cite{ORBBU}. The input channel LLRs ($\bm{\hat{y}}$) are quantized on 5 bits ($Q=5$), with 1 sign bit and 3 bits for the fractional part. Please note that the step-GRAND VLSI architecture has been verified using test benches generated with the bit-true C model of the proposed hardware.

The proposed step-GRAND hardware can support a maximum frequency of $454~\text{MHz}$, and since no pipelining is employed, one time step equals one clock cycle. In the worst-case (W.C.) scenario, the proposed step-GRAND hardware requires $\numprint{279}$ cycles\footnote{ With parameters ($\alpha=2$, $\beta=6$, $P=6$), the $(\gamma,HW)$ values are $(30,3)$,$(18,4)$,$(12,5)$ and $(6,6)$. (Sec. \ref{sec:StairGRANDAlg})} (Eq. (\ref{eq:nb_steps})), which translates to a W.C. latency of 614.5 ns and a W.C. throughput of 170.8 Mbps, as shown in Table \ref{table:StairGRANDPolar}. Figure \ref{fig:lat_tp} illustrates the average latency as well as the average information throughput for the proposed hardware. Please note that the bit-true C model is used to compute the average latency, for the proposed hardware, after taking into account at least 100 frames in error for each $\frac{E_b}{N_0}$ point. As the channel conditions improve, the average latency decreases until it only takes 1 cycle to decode a codeword, enabling up to 47.7 Gbps of information throughput.

As shown in Table \ref{table:StairGRANDPolar}, the proposed step-GRAND hardware implementation requires 35\% less area than the previously proposed ORBGRAND hardware \cite{ORBGRAND-TVLSI}. While the average latency for both ORBGRAND and step-GRAND hardware can be as low as 1 clock cycle (Fig. \ref{fig:lat_tp}), the W.C. latency for ORBGRAND hardware is $\numprint{1918}$ clock cycles \cite{ORBGRAND-TVLSI}\footnote{ Please refer to Fig. 12 of \cite{ORBGRAND-TVLSI} for the W.C. latency of ORBGRAND hardware for various LW and P.}, corresponding to parameters $LW=53$ and $P=6$ (Fig. \ref{fig:FER_polar} (a)), whereas step-GRAND requires only $\numprint{279}$ clock cycles. Therefore, the W.C. latency of step-GRAND hardware $\frac{1}{6.8}\times$ the W.C. latency of ORBGRAND \cite{ORBGRAND-TVLSI}. Furthermore, the step-GRAND is $10\times$ more area-efficient\footnote{$\text{Area Efficiency (Mbps/mm\textsuperscript{2})}=\frac{\text{W.C. Throughput (Mbps)}}{\text{Area (mm\textsuperscript{2})}}$}\footnote{The area efficiency for step-GRAND and  ORBGRAND is 144.7 Mbps/mm\textsuperscript{2} and 13.65 Mbps/mm\textsuperscript{2}, respectively.} than ORBGRAND hardware \cite{ORBGRAND-TVLSI}.

The proposed step-GRAND is also compared with Fixed-Latency (F.L.) ORBGRAND decoder \cite{carloORB} as well as with recent ORBGRAND hardware implementation \cite{ORBBU} that employs a sequential sorter, and the comparison results are shown in Table \ref{table:StairGRANDPolar}. Please note that scaling is not employed due to the vast disparity in technology nodes and implementation methods (Synthesis vs. Integrated Design). The F.L. ORBGRAND \cite{carloORB} deploys $T$ decoder pipeline stages and stores the TEPs ($\bm{e}$) in $T-2$ $Q_s\times{n}$-bit \textit{pattern memories}, where $Q_{max}$ denotes the maximum number of TEPs applied. The proposed step-GRAND reduces the average latency to $2.2$ ns, whereas the F.L. ORBGRAND \cite{carloORB}  offers a fixed latency of $58.49$ ns for decoding polar code (128,105+11). The sequential sorter based ORBGRAND implementation \cite{ORBBU} can achieve an average latency of $40$ ns; the W.C. latency, however, is Not Reported (N.R). The proposed step-GRAND employs a parallel bitonic sorter and, owing to parallel evaluation of TEPs (discussed in section \ref{sec:StairGRANDVLSI}), can achieve a W.C. latency of $614.5$ ns while reducing the average latency to $2.2$ ns. As a result, step-GRAND is suitable for mission-critical applications that have stringent requirements for both average and worst-case latency.


\section{Conclusion}
In this work, we introduce step-GRAND, a soft-input variant of GRAND, along with corresponding hardware architecture. The proposed hardware can decode any linear block code with length $n=128$ and code-rates between $0.75$ and $1$. Furthermore, the step-GRAND introduces parameters that can be tuned to match the desired decoding performance and complexity/latency budget of a target application. The ASIC synthesis results demonstrate that for a code length of $128$ and a target FER of $10^{-7}$, an average information throughput of $47.7$ Gbps can be achieved. Furthermore, compared to the previously proposed ORBGRAND hardware implementation, the proposed step-GRAND hardware is $10\times$ more area efficient and its worst-case latency is $\frac{1}{6.8}\times$ that of the previous ORBGRAND hardware.

\balance
\bibliographystyle{IEEEtran}
\bibliography{IEEEabrv, StepGRAND}
\end{document}